 \documentclass[
twocolumn,showpacs, amsmath, amssymb, floatfix,
nofootinbib,wrapfloat byrevtex]{revtex4}

 \usepackage{amsmath,mathrsfs,graphicx,epstopdf,placeins,float}
 \usepackage{booktabs}
 \usepackage{multirow}
\usepackage{color}

 \newcommand{\beq}[1]{\begin{equation}\label{#1}}
 \newcommand{\eeq}{\end{equation}}
 \newcommand{\bea}[1]{\begin{eqnarray}\label{#1}}
 \newcommand{\eea}{\end{eqnarray}}
 \newcommand\figcaption{\def\@captype{figure}\caption}
 \newcommand\tabcaption{\def\@captype{table}\caption}

\begin{document}

 \title{Asymmetric wave transmission through one dimensional lattices with cubic-quintic nonlinearity}
 \author{Muhammad Abdul Wasay $^{\!\!1,2,3}$}
 \email{wasay31@gmail.com}

\affiliation{$^1$Department of Physics, University of Agriculture, Faisalabad 38040, Pakistan\\
$^2$Center for Photon Information Processing, School of Electrical Engineering and Computer Science, Gwangju Institute of Science and Technology, Gwangju 61005, South Korea\\
$^3$Center for Theoretical Physics of Complex Systems,
Institute for Basic Science (IBS), Daejeon 34051, Republic of Korea}
 \begin{abstract}
One dimensional lattice with an on-site cubic-quintic nonlinear response described by a cubic-quintic discrete nonlinear Schr\"{o}dinger equation is tested for asymmetric wave propagation. The lattice is connected to linear side chains. Asymmetry is introduced by breaking the mirror symmetry of the lattice with respect to the center of the nonlinear region. Three cases corresponding to dimer, trimer and quadrimer are discussed with focus on the corresponding diode-like effect. Transmission coefficients are analytically calculated for left and right moving waves via backward transfer map. The different transmission coefficients for the left and right moving waves impinging the lattice give rise to a diode-like effect which is tested for different variations in asymmetry and site dependent coefficients. We show that there is a higher transmission for incoming waves with lower wavenumbers as compared to the waves with comparatively larger wavenumbers and a diode-like effect improves by increasing the nonlinear layers. We also show that in the context of transport through such lattices, the cooperation between cubic and quintic nonlinear response is not "additive". Finally, we numerically analyse Gaussian wave packet dynamics impinging on the CQDNLS lattice for all three cases.
  \end{abstract}

 \maketitle
 \smallskip

\section{Introduction}

When wave propagates through a medium which is quite sensitive to the incoming wave intensity, one must incorporate some nonlinear corrections to accommodate this when describing wave propagation through such medium. The prospect to model devices (with nonlinear refractive index) that could serve for controlled energy or mass flow is a compelling challenge both from a technological and scientific perspective.

Diode is a device which allows a unidirectional transport and due to this property it can serve as one such device. Search for the diode feature of various phenomenon have been explored in the literature, for example, in acoustics \cite{01,02,03}, heat flow \cite{04,05,06} and electromagnetic waves \cite{07,08,09,010}. When dealing with nonlinear media, the simplest possibility for controlled wave propagation is to devise a 'wave diode' i.e., having asymmetric transmission of waves along two opposite directions of propagation. The hypothesis of reciprocity theorem forbids this possibility in a linear system \cite{011,011a,011b}.

In a linear system, to break the time-reversal symmetry one needs to introduce an external field(electric or magnetic), as is the case in an optical diode. However, much effort can be avoided if we instead have a nonlinear media. This approach \cite{012,013}, seems to be more natural as one can use the nonlinear properties of the media (material) itself to break the parity symmetry and it thus provides a variety of new features for controlled wave propagation.

Nonlinearity leading to asymmetric propagation is studied in many domains. To the best of our knowledge, the first work of this kind in the literature is related to the asymmetric transmission of phonons through a nonlinear layer between two different crystals \cite{014}. The idea has been popular in the field of nonlinear optics, for example in \cite{015,016}, a so called all-optical diode was
proposed. In \cite{017}, a thermal diode was proposed which is capable of transmitting heat asymmetrically between two different sources by means of nonlinearity, a similar study in \cite{018} was done for the rectification of heat conduction by means of the asymmetry and non-harmonic nature of the system.
 The propagation of sound waves is another important issue and a so-called acoustic wave diode was proposed in \cite{019}.

In a discrete nonlinear setting, wave propagation has been studied in different physical contexts \cite{020,021,022,023}. The discreteness in such systems is attributed to a weak interaction between different elements of the system, for instance, BEC trapped in optical lattices, and coupled optical waveguides.

In the context of asymmetric wave transmission through a 1D layered photonic crystal lattice, it will be interesting to see how transmission is effected by a higher order (quintic) on-site nonlinear response.  We will use a set of discrete nonlinear Schr\"{o}dinger (DNLS) equations with a local (on-site) cubic-quintic nonlinearity. This cubic-quintic DNLS model has been studied for mobility regimes of solitons in 1D lattices \cite{024} and in 2D lattices \cite{025}. DNLS models have also been used to study various related phenomenon \cite{26,27,28}.

In this paper we will work with a DNLS equation having on-site focusing cubic-quintic nonlinearity. We will investigate the scattering phenomenon for two, three and four nonlinear layers. The DNLS equation is equipped with variable site dependant coefficients in order to describe the nonlinear features of different layers. The system is such that these nonlinear sites are embedded into a linear lattice, and are connected to linear side chains. As a physically relevant case we will also analyse the dynamics of an incident Gaussian wave packet on this CQDNLS lattice system for all three cases.


\section{The Model}

The set up assumes an on-site (local) cubic-quintic nonlinear response, which can be modelled by a set of discrete nonlinear Schr\"{o}dinger equations with local cubic-quintic nonlinearity.

The time dependent DNLS with focusing cubic-quinic nonlinearity is given by\cite{024}\cite{025}

\bea{}
\!i\frac{d\psi_n}{dt}\!=\!V_n\psi_n\!-\!(\psi_{n+1}+\psi_{n-1})\!+\!\gamma_n|\psi_n|^2\psi_n\!+\!\nu_n|\psi_n|^4\psi_n
\label{first}
\eea

Here $V_n$ is the potential on site $n$, $\gamma_n$ and $\nu_n$ represent the on-site cubic and quintic nonlinearities respectively, and $\gamma_n,\nu_n>0$.  $\psi_n$ is the amplitude of the field at the $n$-th lattice site. The Hamiltonian is

\bea{}
H=-\sum_n[(\psi_n^*\psi_{n+1}+\psi_n\psi_{n+1}^*)+V_n|\psi_n|^2+\nonumber
\\
\frac{\gamma_n}{2}|\psi_n|^4+\frac{\nu_n}{3}|\psi_n|^6]
\eea

with this Hamiltonian one can derive the equation of motion \eqref{first} by using

\bea{}
i\frac{d{\psi}_n}{dt}=\frac{\partial H}{\partial \psi_n^*}
\eea

 The dynamical equations \eqref{first} have solutions of the form $\psi(t)=\phi e^{-i\omega t}$, where $\phi$ is independent of $t$, substituting this in \eqref{first} leads to

\bea{}
\omega\phi_n=V_n\phi_n-\phi_{n+1}-\phi_{n-1}+\gamma_n|\phi_n|^2\phi+\nu_n|\phi_n|^4\phi_n
\label{second}
\eea

where $\omega$ is the spatial frequency and $\phi_n$ is the complex amplitude on site $n$ with potential $V_n$. The nonlinear sites are embedded in a linear lattice and are connected to linear side chains where the wave can propagate freely, therefore we can say that $\gamma_n,\nu_n$ and $V_n$ are non-vanishing only for $1\leq n\leq N$, here $n$ represents nonlinearity at a particular site and $N$ represents the total number of nonlinear sites. Let us consider plane wave solutions of the form

\bea{}
\phi_n=
\left(
  \begin{array}{c}
  R_0 e^{ikn}+Re^{-ikn}             \qquad n\leq 1\\
  Te^{ikn}                          \qquad\qquad\qquad n\geq N \\
  \end{array}
\right)
\eea

where $R_0,R$ and $T$ are the amplitudes of incident, reflected and transmitted wave respectively. As mentioned above, in the linear region with $n>N$ or $n<1$,  the wave propagates freely. The spatial frequency is $\omega=-2~\textmd{cos}(k)$.

For $n=0$, we have
\bea{}
\phi_0=R_0+R
\eea
and for $n=1$
\bea{}
\phi_1=R_0e^{ik}+Re^{-ik}
\eea
With $\phi_0$ and $\phi_1$ we can get $R$ and $R_0$ in terms of $\phi_0$ and $\phi_1$
\bea{}
R=\frac{\phi_0 e^{ik}-\phi_1}{e^{ik}-e^{-ik}}
\label{3rd}
\eea
and
\bea{}
R_0=\frac{\phi_0 e^{-ik}-\phi_1}{e^{-ik}-e^{ik}}
\label{4th}
\eea

The extent to which an incident wave is transmitted is calculated by the transmission coefficient $t(k,|T|^2)=\frac{|T|^2}{|R_0|^2}$. We will calculate these coefficients by a backward transfer map \cite{026,027,028}, obtained by rearranging Eq.\eqref{second},

\bea{}
\phi_{n-1}=-\phi_{n+1}+(V_n-\omega+\gamma_n|\phi_n|^2+\nu_n|\phi_n|^4)\phi_n
\label{5th}
\eea

It is useful to introduce the following notation for later convenience, we define

\bea{}
\delta_j=V_j-\omega+\gamma_j|\phi_j|^2+\nu_j|\phi_j|^4
\label{6th}
\eea

\section{Results}

With the backward iterative map at hand, we will now compute the transmission coefficients for the case of dimer (two nonlinear sites), trimer (three nonlinear sites) and quadrimer (four nonlinear sites) in the following subsections.

\subsection{DIMER}

 We consider the simplest case of two nonlinear layers, i.e., $N=2$: The dimer. The cubic-quintic DNLS with $N=3$ and $N=4$ will be considered in the following subsections. From Eq.\eqref{5th} with $n=2$,

\bea{}
\phi_1=Te^{2ik}(\delta_2-e^{ik})
\eea

where, $\delta_2=(V_2-\omega+\gamma_2|T|^2+\nu_2|T|^4)$

For $n=1$, we have

\bea{}
\phi_0=Te^{2ik}[\delta_1(\delta_2-e^{ik})-1]
\eea
where, $\delta_1=V_1-\omega+\gamma_1|T|^2|(\delta_2-e^{ik})|^2+\nu_1|T|^4|(\delta_2-e^{ik})|^4$\\

and from Eq.\eqref{4th}

\bea{}
|R_0|^2=\frac{|T|^2|(\delta_2-e^{ik})(\delta_1-e^{ik})-1|^2}{|e^{-ik}-e^{ik}|^2}
\eea
\bea{}
 \Rightarrow |T|^2=\frac{|R_0|^2|e^{-ik}-e^{ik}|^2}{|(\delta_2-e^{ik})(\delta_1-e^{ik})-1|^2}
\eea

So, the sought transmission coefficient is

\bea{}
t(k,|T|^2)=\left|\frac{e^{-ik}-e^{ik}}{(\delta_2-e^{ik})(\delta_1-e^{ik})-1}\right|^2
\label{7th}
\eea

We have calculated the transmission coefficient for left moving wave with $k>0$. To calculate the transmission coefficient for right moving wave
we assume that the sample is flipped such that the encountered sites are now labelled as,$1\rightarrow N',2\rightarrow (N-1)',...,N\rightarrow 1$ so that
 to solve for transmission coefficient with $k<0$ we simply need to change $V_n=V_{N-n+1}$ i.e to change $V_1$ by $V_2$ and vice versa. Let us introduce the notation


\bea{}
\zeta_2=V_1-\omega+\gamma_1|T'|^2+\nu_1|T'|^4
\eea

and

\bea{}
\!\!\!\!\!\zeta_1\!=\!V_2-\omega\!+\!\gamma_2|T'|^2|(\zeta_2-e^{ik'})|^2\!+\!\nu_2|T'|^4|(\zeta_2-e^{ik'})|^4
\eea

Using a backward transfer map as discussed above, for the left moving wave with $k'=-k$ we get

\bea{}
|R'_0|^2=\frac{|T'|^2|(\zeta_2-e^{ik'})(\zeta_1-e^{ik'})-1|^2}{|e^{-ik'}-e^{ik'}|^2}
\eea
and thus the transmission coefficient turns out to be

\bea{}
t'(k',|T'|^2)=\left|\frac{e^{-ik'}-e^{ik'}}{(\zeta_2-e^{ik'})(\zeta_1-e^{ik'})-1}\right|^2
\label{8th}
\eea

The lattice is still mirror symmetric with respect to the center of the nonlinear region, we must break this symmetry to achieve an asymmetric transmission leading to the desired diode effect. Once the symmetry is broken, we will have different transmission coefficients for left and right moving wave. This can be done in different ways. We break the symmetry by chosing different potentials on each of the two nonlinear sites, i.e., $V_n \not= V_{N-n+1}$. For dimer $N=2$, we take $V_n=V_0(1\pm\varepsilon)$, and we choose to have positive sign at site 1 and negative sign at site 2, $\varepsilon$ is the extent of asymmetry and $V_0$ is the depth of the potential.

\begin{figure}[htb]
  \centering
  \includegraphics[scale=0.33]{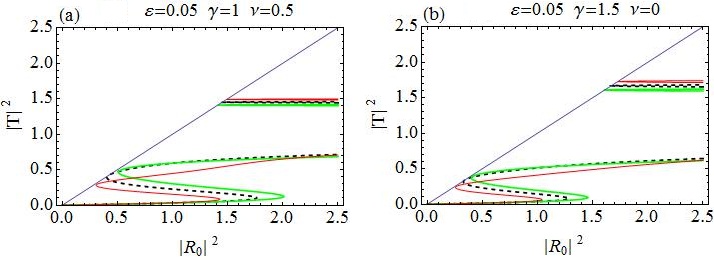}
  \caption{$|T|^2$ as function of $|R_0|^2$. (a): \textit{CQDNLS dimer}. (b): \textit{Purely cubic DNLS dimer}}
\end{figure}

Fig.$1$ depicts the relationship between incident ($|R_0|^2$ along horizontal axis) and transmitted intensity ($|T|^2$ along vertical axis). Transmission curves in Fig.1(a) correspond to CQDNLS dimer with $V_0=-2.5,\varepsilon=0.05,\gamma=1,\nu=0.5$ and $|k|=0.1$, and in Fig.1(b) for a purely cubic DNLS dimer with $V_0=-2.5,\varepsilon=0.05,\gamma=1.5$ and $|k|=0.1$. The dashed black line corresponds to the symmetric case $\varepsilon=0$. The asymmetric branch with $\varepsilon\not=0$ have two oppositely directed waves with differently detuned resonances responsible for different transmission coefficients and thus for the asymmetric transmission. The first window of maximal transmission in CQDNLS model is slightly broader as compared to the corresponding window in cubic DNLS model \cite{012,030} with same parameter values, however, this trend is reversed for the second window in both models. Moreover, the first windows occur roughly at the same incoming intensities in both models, while the second window is displaced to slightly lower intensities in the CQDNLS model as compared to the cubic DNLS model. Further, please note that the bistable behavior persists in the CQDNLS dimer model, which is apparent from the corresponding (bistability) windows in Fig.1(a).

The purpose of plotting curves for the cubic DNLS dimer stems from the idea that one could possibly mimic similar transmission curves (as for the CQDNLS dimer) by increasing the purely cubic response in a cubic DNLS dimer. However, as one can see in Fig.1, the transmission pattern is somewhat different in Fig.1(a) and Fig.1(b), with same parameters except we chose $\gamma=1.5$ instead of $\gamma=1$ since we enhanced the cubic response in the hope that this could lead to the same transmission curves as for the CQDNLS model. Thus the cooperation between cubic and quintic response in a CQDNLS model does not seem to be of an ''additive" type which one usually presumes.

Further also note that in the CQDNLS dimer case, the windows of maximal transmission keep broadening with the corresponding increase in asymmetry level until $\varepsilon\sim 0.1$, where the first window starts to diminish.


\begin{figure}[htb]
  \centering
  \includegraphics[scale=0.4]{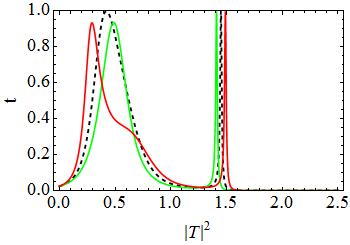}
  \caption{Transmission coefficient $t$ as a function of transmitted intensity $|T|^2$: \textit{For Dimer}}
\end{figure}
\vspace{-3mm}

Fig.$2$ shows the transmission coefficient $t(k,|T|^2)$ along vertical axis as a function of transmitted intensity $|T|^2$ along horizontal axis with $V_{1,2}=V_0(1\pm\varepsilon),N=2,V_0=-2.5,\gamma=1,\nu=0.5,\varepsilon=0.05,|k|=0.1$

\subsubsection{Transmission Coefficient}

In this subsection we present plots for the transmission coefficient as a function of transmitted intensity $|T|^2$ and $k$. Fig.3 corresponds to dimer with varying asymmetry while the quintic response fixed at $\nu=0.5$, and all other parameters as before. The transmission seems to reduce as we increase the asymmetry, specially above $\varepsilon=0.2$.
\begin{figure}[h]
  \centering
  \includegraphics[scale=0.38]{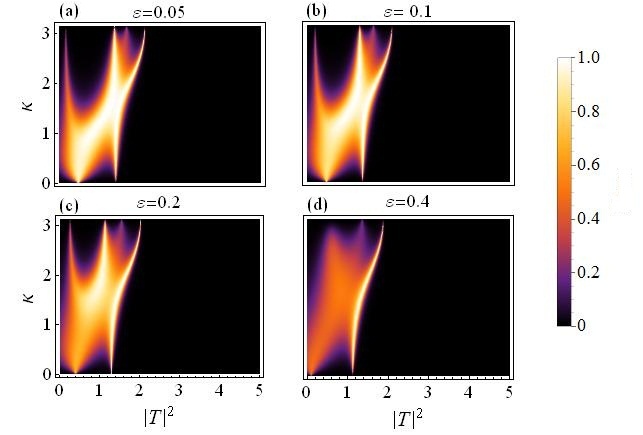}
  \caption{Transmission coefficient as a function of $|T|^2$ and $k$. Varying asymmetry level:(a) $\varepsilon=0.05$  to (d) $\varepsilon=0.4$}
\end{figure}

Fig.4 is produced by fixing the asymmetry to $\varepsilon=0.05$ and letting $\nu$ (the quintic nonlinearity) strengthen as compared to $\gamma(=1)$, the cubic nonlinearity. The plot in Fig.4(a) corresponds to the purely cubic case ($\nu=0$), while we increase $\nu$ in the subsequent plots from $\nu=0.3$ in Fig.4(b) to $\nu=0.7$ in Fig.4(d). As $\nu$ is increased, the overall transmission is reduced, however, this effect is not significantly large for waves with small $k$, as compared to the waves with large $k$. Note that with asymmetry fixed at $\varepsilon=0.05$ the region for maximal transmission with CQ nonlinearity is smaller as compared to the purely cubic case \cite{012}.

\begin{figure}[h]
  \centering
  \includegraphics[scale=0.38]{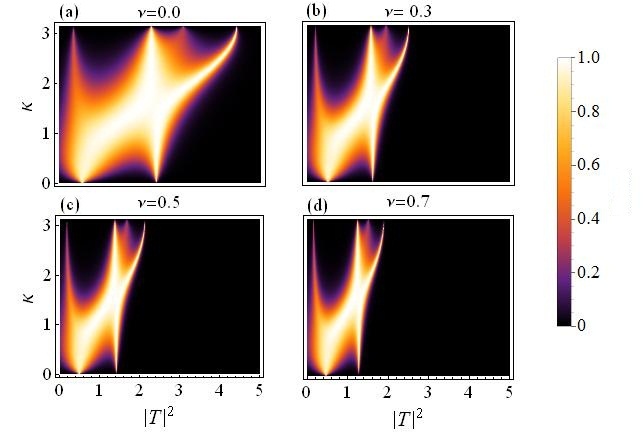}
  \caption{Transmission coefficient as a function of $|T|^2$ and $k$: \textit{For varying quintic nonlinear response}.}
\end{figure}

Finally, in both figures Fig.3 and Fig.4, one may note that there are two transmission peaks which split into four as $k$ increases. This split occurs roughly around $k\sim\pi/2$.

Fig.5 below corresponds to the same parameter strengths as in Fig.1, and is plotted to highlight the type of cooperation between cubic and quintic nonlinearities. It is evident that a purely cubic DNLS model allows a higher transmission of the input waves as compared to a cubic-quintic DNLS model with the same strength of the nonlinear response distributed between cubic and quintic order.

\begin{figure}[h]
  \centering
  \includegraphics[scale=0.38]{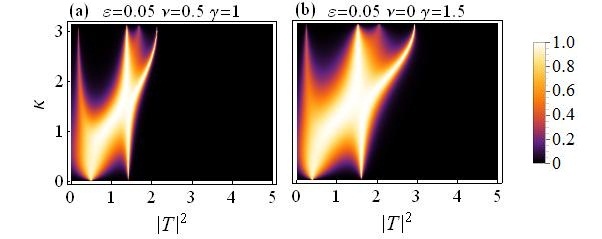}
  \caption{Transmission coefficient as a function of $|T|^2$ and $k$: (a): CQDNLS dimer with the nonlinear strength distributed between cubic and quintic response. (b): Cubic DNLS dimer with the same nonlinear strength purely as cubic response}
\end{figure}

\subsubsection{Rectifying Factor}

To see where the best diode effect occurs, following \cite{012}, a rectifying factor is introduced as follows

\bea{}
f=\frac{t(k,|T|^2)-t(-k,|T|^2)}{t(k,|T|^2)+t(-k,|T|^2)}
\label{Rfactor}
\eea

Fig.6 is a plot for rectifying factor as a function of $|T|^2$ and $k$, for the dimer case.
\begin{figure}[h]
  \centering
  \includegraphics[scale=0.38]{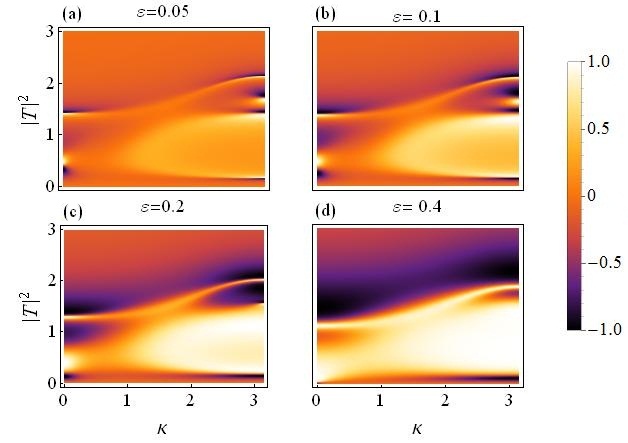}
  \caption{Rectifying factor as a function of $|T|^2$ and $k$: \textit{For varying asymmetry level.}}
\end{figure}

The asymmetry is increased from $\varepsilon=0.05$ in (a) to $\varepsilon=0.4$ in (d), the change in color indicates improved diode effect with increasing asymmetry, but a reduced overall transmission.

\subsubsection{Gaussian wavepacket dynamics for dimer}

It is instructive to look at a Gaussian wave packet scattering for CQDNLS lattice system. We consider a dimer embedded inside a lattice with $M=1000$ sites with open boundary conditions. The dimer is placed at 500th and 501th site. The wave packet at $t=0$ (initial condition) for the right incidence is of the form \cite{012,013}

\bea{}
\psi_n(0)=B ~\textmd{exp}\left[\frac{-(n-n_0-\frac{M}{2})^2}{w^2}+ik_0(n-\frac{M}{2})\right]
\label{GWP1}
\eea

$B$ is the incoming amplitude, $n_0$ lattice starting point, $w$ width of wavepacket.

The initial wave packet for the left incidence is given by ($k_0\rightarrow-k_0$)

\bea{}
\psi_n(0)\!=\!B ~\textmd{exp}\left[\!\frac{-(\!n\!-\!n_0\!-\!\frac{M}{2}\!-M\!)^2}{w^2}\!-\!ik_0(\!n\!-\!\frac{M}{2}\!-\!M\!)\right]
\label{GWP2}
\eea

\begin{figure}[h]
  \centering
  \includegraphics[scale=0.30]{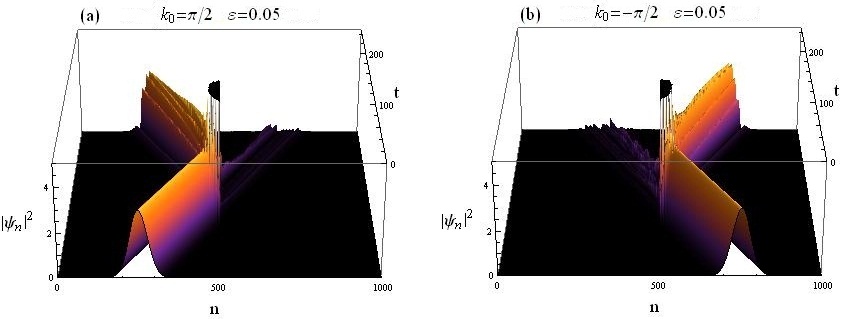}
  \caption{Gaussian wavepacket impinging on CQDNLS dimer. (a): Right incidence (b): Left incidence}
\end{figure}

  Fig.7 depicts how the wave packet is scattered when it hits the CQDNLS dimer in the middle of the lattice.
For the same system parameters, the wave packet transmission coefficients are found to be $t_{k>0}=0.0761197$ for right moving packet and $t_{k<0}=0.116237$ for left moving packet.

\subsection{TRIMER}
\vspace{-2mm}
We want to examine the transmission phenomenon for the case when we have three nonlinear layers (sites), i.e., a trimer, $N=3$. Adopting the same procedure of backward transformer map as above, the transmission coefficient for trimer is found to be

\bea{}
t(k,|T|^2)=\left|\frac{e^{ik}-e^{-ik}}{e^{ik}-\delta_1+(e^{ik}-\delta_3)(1-\delta_2(\delta_1-e^{ik}))}\right|^2
\eea
with
\bea{}
\delta_3=V_3-\omega+\gamma_3|T|^2+\nu_3|T|^4
\eea
\bea{}
\delta_2=V_2-\omega+\gamma_2|T|^2|\delta_3-e^{ik}|^2+\nu_2|T|^4|\delta_3-e^{ik}|^4
\eea
\bea{}
\delta_1=V_1-\omega+\gamma_1|T|^2|\delta_2(\delta_3-e^{ik})-1|^2+\nonumber
\\
\nu_1|T|^4|\delta_2(\delta_3-e^{ik})-1|^4
\eea

\begin{figure}[h]
  \centering
  \includegraphics[scale=0.33]{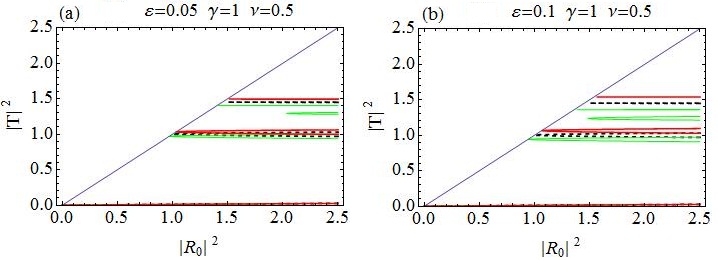}
  \caption{$|T|^2$ as function of $|R_0|^2$. (a): CQDNLS Trimer with $\varepsilon=0.05$. (b): CQDNLS Trimer with $\varepsilon=0.1$}
\end{figure}


Plots in Fig.8 represent relationship between incident intensity $|R_0|^2$ and transmitted intensity $|T|^2$. The maximal transmission occurs at two intervals. Plot in Fig.8(b) corresponds to an increased level of asymmetry $\varepsilon=0.1$ with all other parameters as before, the intervals of maximal transmission broaden, but around $\varepsilon\sim 0.16$ only first window for maximal transmission corresponding to lower incoming intensity survives, while the window corresponding to the higher incoming intensity diminishes above a critical value of asymmetry $\varepsilon\sim 0.16$. This is in contrast to the dimer case where instead the first diode window diminishes.

\begin{figure}[h]
  \centering
  \includegraphics[scale=0.4]{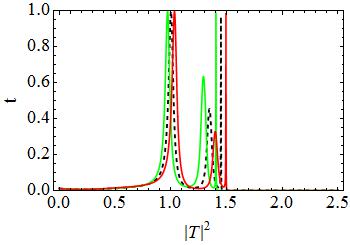}
  \caption{Transmission coefficient $t$ as a function of transmitted intensity $|T|^2$: \textit{For Trimer}}
\end{figure}

We find that the first window for maximal transmission has been displaced to higher incoming intensities and both windows have shrunk (for fixed asymmetry) as compared to the dimer case. However, increasing asymmetry can broaden the windows upto ($\varepsilon\sim 0.16$) after which only one diode window survives. The overall transmission seems to be reduced as compared to the dimer.

Fig.9 shows for trimer, the transmission coefficient $t(k,|T|^2)$ along vertical axis as a function of transmitted intensity $|T|^2$ along horizontal axis with $V_{1,3}=V_0(1\pm\varepsilon),N=3,,V_2=V_0,V_0=-2.5,\gamma=1,\nu=0.5,\varepsilon=0.05,|k|=0.1$.
\vspace{-5mm}

\subsubsection{Transmission Coefficient}
\vspace{-3mm}
As before, transmission coefficient as a function of transmitted intensity $|T|^2$ and $k$ for increasing asymmetry level for the trimer are plotted in Fig.10 below. The transmission peak splitting phenomenon occurs again roughly around $k=\pi/2$. Waves with smaller $k(\sim\leq\pi/2)$ have a higher transmission than for the waves with larger $k(\sim\geq\pi/2)$. Overall transmission is reduced as asymmetry is increased from $\varepsilon=0.05$ in (a) to $\varepsilon=0.4$ in (d), and all other parameters as before.

\begin{figure}[h]
  \centering
  \includegraphics[scale=0.38]{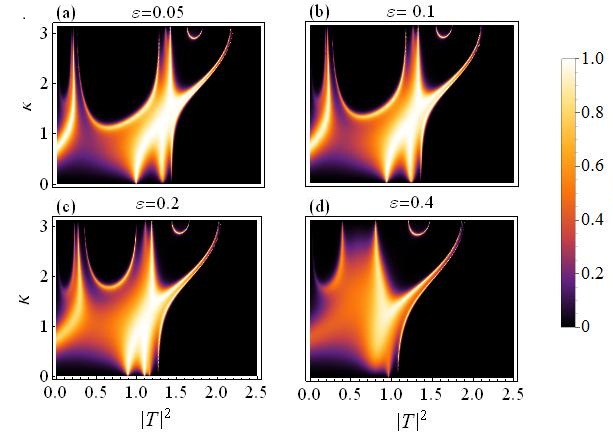}
  \caption{Transmission coefficient as a function of $|T|^2$ and $k$. For varying asymmetry level (Trimer). (a): $\varepsilon=0.05$ to (d): $\varepsilon=0.4$}
\end{figure}
\subsubsection{Rectifying Factor}
\vspace{-3mm}
Using \eqref{Rfactor}, the rectifying plots in Fig.11 for trimer case with varying asymmetry $\varepsilon=0.05$ to $\varepsilon=0.4$ are given below. Diode-like action seems better than the dimer case and improves with increasing asymmetry but the overall transmission is reduced as compared to the dimer.

\begin{figure}[h]
  \centering
  \includegraphics[scale=0.38]{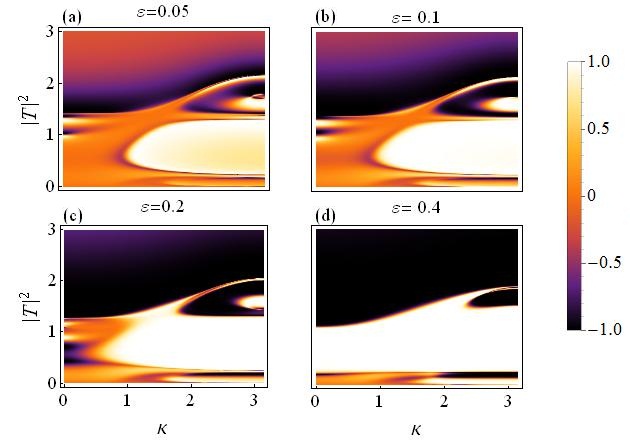}
  \caption{Rectifying factor as a function of $|T|^2$ and $k$: For varying asymmetry level (Trimer). (a): $\varepsilon=0.05$ to (d): $\varepsilon=0.4$}
\end{figure}
\vspace{-6mm}
\subsubsection{Gaussian wavepacket dynamics for trimer}

The scattering of an incoming Gaussian wave packet by a CQDNLS trimer in the middle of the lattice system is depicted in Fig.12 below. We take Eqs. \eqref{GWP1},\eqref{GWP2} as the initial condition. The trimer in this case corresponds to sites 500, 501 and 502 in the lattice with 1000 sites.

The on site potential is chosen as $V_{500,502}=V_0(1\pm\varepsilon)$ and $V_{501}=V_0$, remaining parameters as before.

\begin{figure}[h]
  \centering
  \includegraphics[scale=0.30]{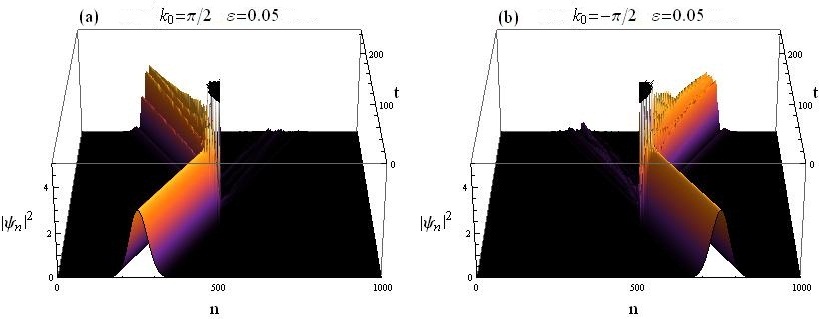}
  \caption{Gaussian wavepacket impinging on CQDNLS Trimer. (a): Right incidence. (b): Left incidence.}
  \end{figure}


The wave packet transmission coefficients for both left and right incidences are: $t_{k>0}=0.0301173$ for right moving packet and $t_{k<0}=0.0645591$ for left moving packet.

\subsection{QUADRIMER}

We keep on increasing the number of nonlinear layers, and now we examine transmission phenomenon through four nonlinear sites, i.e., a quadrimer.
For quadrimer then $N=4$. Using backward transformer map, the transmission coefficient for quadrimer is found to be

\bea{}
|R_0|^2=~~~~~~~~~~~~~~~~~~~~~~~~~~~~~~~~~~~~~~~~~~~~~~~~~~~~~~~~~~~~~~~~~~
\\
\hspace{-10mm}\frac{|T|^2|1+\delta_2(e^{ik}\!-\!\delta_1)+(\delta_4\!-\!e^{ik})(\delta_1\delta_2\delta_3\!-\!e^{ik}\delta_2\delta_3\!-\!\delta_3\!-\!\delta_1+e^{ik})|^2}{|e^{ik}-e^{-ik}|^2}\nonumber
\eea

\bea{}
\hspace{-10mm}t(k,|T|^2)=~~~~~~~~~~~~~~~~~~~~~~~~~~~~~~~~~~~~~~~~~~~~~~~~~~~~~~~~~~~~~~~
\\
\hspace{-10mm}\left|\frac{e^{ik}-e^{-ik}}{1+\delta_2(e^{ik}-\delta_1)+(\delta_4-e^{ik})(\delta_1\delta_2\delta_3-e^{ik}\delta_2\delta_3-\delta_3-\delta_1+e^{ik})}\right|^2\nonumber
\eea
with,
\bea{}
\delta_4=V_4-\omega+\gamma_4|T|^2+\nu_4|T|^4
\eea
\bea{}
\delta_3=V_3-\omega+\gamma_3|T|^2|\delta_4-e^{ik}|^2+\nu_3|T|^4|\delta_4-e^{ik}|^4
\eea
\bea{}
\delta_2=V_2-\omega+\gamma_2|T|^2|\delta_3(\delta_4-e^{ik})-1|^2+
\\
\nu_2|T|^4|\delta_3(\delta_4-e^{ik})-1|^4\nonumber
\eea
\bea{}
\delta_1=V_1-\omega+\gamma_1|T|^2|(\delta_4-e^{ik})(\delta_2\delta_3-1)-\delta_2|^2+
\\
\nu_1|T|^4|(\delta_4-e^{ik})(\delta_2\delta_3-1)-\delta_2|^4\nonumber
\eea

\begin{figure}[htb]
  \centering
  \includegraphics[scale=0.4]{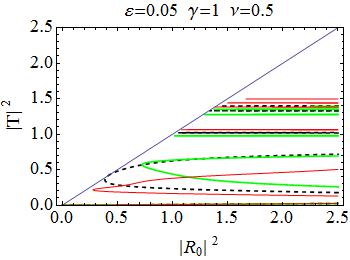}
  \caption{$|T|^2$ as function of $|R_0|^2$: \textit{For Quadrimer}}
\end{figure}

Fig.$13$ represents $|T|^2$ as function of $|R_0|^2$, for $N=4$. there is no diode effect for smaller incoming intensities $\sim |R_0|^2\leq1$, however there are high transmission peaks for higher incoming intensities, which become complicated with different merging peaks.

\begin{figure}[h]
  \centering
  \includegraphics[scale=0.4]{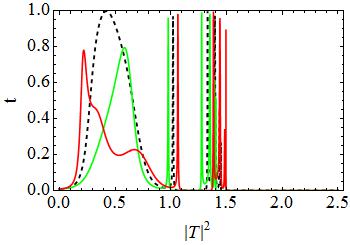}
  \caption{Transmission coefficient $t$ as a function of transmitted intensity $|T|^2$: \textit{For Quadrimer}}
\end{figure}

Fig.$14$ is for $t(k,|T|^2)$ vs $|T|^2$, with parameter values: $V_{1,4}=V_0(1\pm\varepsilon),N=4,V_2=V_3=V_0,V_0=-2.5,\gamma=1,\nu=0.5,\varepsilon=0.05$ and $|k|=0.1$.

We can see that as $N$ increases, the transmission pattern becomes complicated and it becomes difficult to identify the peak shift which occurs due to asymmetry. As compared to the simplest case of two sites, the transmission pattern for increased sites is rather complicated and regions with maximal transmission are scarce, see \cite{029} for a related discussion on a cubic DNLS model.

\subsubsection{Transmission Coefficient}

The transmission coefficient is plotted in Fig.15 for the quadrimer. There seems to be very little transmission for larger wavenumbers $k$, as compared to a slightly better transmission for smaller wavenumbers. The transmission tends to reduce as we increase the asymmetry. The overall transmission for quadrimer is smaller than both the dimer and trimer case.

\begin{figure}[h]
  \centering
  \includegraphics[scale=0.38]{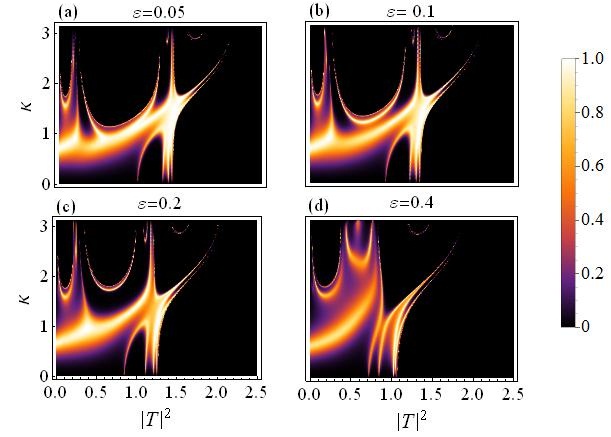}
  \caption{Transmission coefficient as a function of $|T|^2$ and $k$: \textit{For varying asymmetry level (Quadrimer)}.}
\end{figure}
\vspace{-5mm}
\subsubsection{Rectifying Factor}

 Rectifying factor for $N=4$ with increasing asymmetry level is presented in Fig.16 below. The diode effect improves but with overall transmission smaller than the preceding cases (dimer and trimer).
\begin{figure}[h]
  \centering
  \includegraphics[scale=0.38]{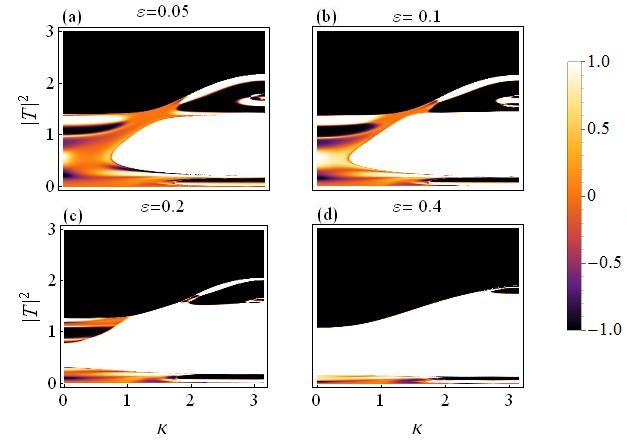}
  \caption{Rectifying factor as a function of $|T|^2$ and $k$: \textit{For varying asymmetry level (Quadrimer)}.}
\end{figure}
\vspace{20mm}
\subsubsection{Gaussian wavepacket dynamics for quadrimer}

How an incoming Gaussian wave packet specified by the initial conditions given in Eqs. \eqref{GWP1} and \eqref{GWP2} is scattered by a CQDNLS quadrimer placed at the center of the lattice, is shown in Fig.17. The quadrimer is chosen to be embedded at sites 500,501,502 and 503. Along the lines of our previous discussion on quadrimer, we chose on site potentials as: $V_{500,503}=V_0(1-\varepsilon)$ and $V_{501,502}=V_0$, other parameters as above.

\begin{figure}[h]
  \centering
  \includegraphics[scale=0.30]{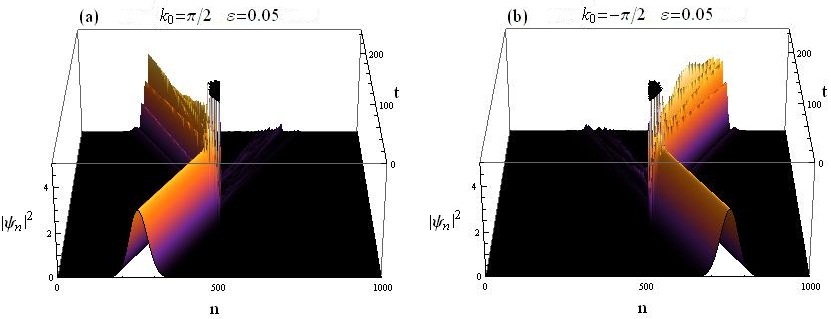}
  \caption{Gaussian wavepacket impinging on CQDNLS Quadrimer. (a): Right incidence. (b): Left incidence.}
  \end{figure}


Transmission coefficients for the wave packet in this case are: $t_{k>0}=0.0558794$ transmission coefficient for right moving packet and $t_{k<0}=0.0545692$ coefficient for left moving packet.

\section{SUMMARY AND CONCLUSION}

We have investigated wave propagation through a nonlinear system having an on-site cubic-quintic nonlinear response, using a set of cubic-quintic discrete nonlinear Schr\"{o}dinger equations with site dependent coefficients as a model of the system. The purpose of this paper was to examine this model for scattering phenomenon and as a wave diode candidate. We introduced an asymmetry in the form of different on-site potentials to break the mirror symmetry of the system which resulted in an asymmetric transmission of the incoming waves. Bistable behavior leading to an asymmetric transmission also persists in all three cases, i.e., dimer, trimer and quadrimer.

Based on the presumption that a slightly lower transmission in CQDNLS as compared to its purely cubic DNLS counterpart \cite{012} could be due to an "additive cooperation" between cubic and quintic response, we tested the dimer and it turned out that this cooperation is not of an "additive" type. A more exhaustive investigation into the "type of this cooperation" is left for the future.

We probed for parameter regimes where this CQDNLS model can be used as a wave diode. We examined the cases of a dimer, trimer and a quadrimer and found that when we increase the nonlinear sites from two, the transmission pattern becomes complicated while retaining some diode behavior which improves with increasing asymmetry but smaller overall transmission, see \cite{012,029} and references therein.

Another important aspect of this study is that one of the windows of maximal transmission diminish after reaching a critical value of asymmetry, for dimer and trimer. After $\varepsilon\sim 0.1$, first window disappears for the dimer, whereas in trimer case the second window disappears after $\varepsilon\sim 0.16$.

In the trimer case a diode-like effect occurs at higher incoming intensities as compared to the dimer, and for the quadrimer, the transmission pattern gets too complicated. Moreover, increasing the strength of the on-site quintic nonlinear response was tested for the dimer case. It results in lower transmission on the whole, but for waves with small wavenumber, the maximal transmission does not seem to be effected much as compared to the waves with larger wavenumber. This pattern carries on for the trimer and the quadrimer case with fixed quintic nonlinear response ($\nu=0.5$). It is important to note that in contrast to the trimer and quadrimer, in the dimer case this phenomenon occurs by increasing the on-site quintic response, not the asymmetry. We also presented plots for the rectifying factor for dimer, trimer and quadrimer to present the corresponding diode effect and we can say that the diode effect is enhanced with increasing asymmetry as we increase the number of nonlinear layers at the cost of an overall lower transmission.

We also observed a transmission peak splitting pattern which roughly occurs around $k\sim\frac{\pi}{2}$ in the plots for transmission coefficient for the dimer, trimer and quadrimer.

Finally, to put the theoretical work in context with physically relevant phenomenon, we presented results of our numerical considerations of how a Gaussian wave packet impinging on the CQDNLS lattice system is scattered by a dimer, trimer and quadrimer with their corresponding transmission coefficients.

As a future prospect, it will be interesting to see what happens when we saturate the nonlinear response \cite{031}, then we have an extra parameter to play with and perhaps we can fine tune the model and test it for better transmission. The work in this direction is in progress and we will soon report our results on this.

\section*{Acknowledgments}

 M.A. Wasay would like to thank the IBS Center for Theoretical Physics of Complex system in Daejeon, South Korea for hospitality and financial support where initial part of this work was carried out.

\section*{Competing Financial Interests}

The author declares no competing interests.


\begin{thebibliography}{99}

\expandafter\ifx\csname natexlab\endcsname\relax\def\natexlab#1{#1}\fi
\expandafter\ifx\csname bibnamefont\endcsname\relax
  \def\bibnamefont#1{#1}\fi
\expandafter\ifx\csname bibfnamefont\endcsname\relax
  \def\bibfnamefont#1{#1}\fi
\expandafter\ifx\csname citenamefont\endcsname\relax
  \def\citenamefont#1{#1}\fi
\expandafter\ifx\csname url\endcsname\relax
  \def\url#1{\texttt{#1}}\fi
\expandafter\ifx\csname urlprefix\endcsname\relax\def\urlprefix{URL }\fi
\providecommand{\bibinfo}[2]{#2}
\providecommand{\eprint}[2][]{\url{#2}}



\bibitem{01} Li, X. F. \textit{et al.} Tunable Unidirectional Sound Propagation through a Sonic-Crystal-Based Acoustic Diode. \textit{Phys. Rev. Lett.} \textbf{106}, 084301 (2011).
\bibitem{02} Boechler, N., Theocharis, G., and Daraio, C. Bifurcation-based acoustic switching and rectification. \textit{Nat. Mater.} \textbf{10}, 665-668 (2011).
\bibitem{03} Yuan, B., Liang, B., Tao, J. C., Zou, X. Y., and Cheng, J. C. Broadband directional acoustic waveguide with high efficiency. \textit{Appl. Phys. Lett.} \textbf{101}, 043503 (2012).
\bibitem{04} Chang, C. W., Okawa, D., Majumdar, A. and Zettl, A. Solid-State Thermal Rectifier. \textit{Science} \textbf{314}, 1121-1124 (2006).
\bibitem{05} Sun, T., Wang, J. X., and Kang, W. Ubiquitous thermal rectification induced by non-diffusive weak scattering at low temperature in one-dimensional materials: Revealed with a non-reflective heat reservoir. \textit{Europhys. Lett.} \textbf{105}, 16004 (2014).
\bibitem{06} Wang, Y. , Vallabhaneni, A., Hu, J. N., Qiu, B., Chen, Y. P., and Ruan, X. L. Phonon Lateral Confinement Enables Thermal Rectification in Asymmetric Single-Material Nanostructures. \textit{Nano Lett.} \textbf{14}, 592-596 (2014).
\bibitem{07} Gallo, K., Assanto, G., Parameswaran, K. R., and Fejer, M. M. All-optical diode in a periodically poled lithium niobate waveguide. \textit{Appl. Phys. Lett.} \textbf{79}, 314 (2001).
\bibitem{08} Fan, L. \textit{et al.} An All-Silicon Passive Optical Diode. \textit{Science} \textbf{335}, 447-450 (2012).
\bibitem{09} Roy, D. Few-photon optical diode. \textit{Phys. Rev. B} \textbf{81}, 155117 (2010).
\bibitem{010} Lira, H., Yu, Z. F., Fan, S. H., and Lipson, M. Electrically Driven Nonreciprocity Induced by Interband Photonic Transition on a Silicon Chip. \textit{Phys. Rev. Lett.} \textbf{109}, 033901 (2012).
\bibitem{011} Rayleigh, J. The Theory of Sound. \textit{Dover Publications}, New York, (1945).
\bibitem{011a} Figotin, A.  and Vitebsky, I. Nonreciprocal magnetic photonic crystals. \textit{Phys. Rev. E} \textbf{63}, 066609 (2001).
\bibitem{011b} Khanikaev, A. B. and Steel, M. J. Low-symmetry magnetic photonic crystals for nonreciprocal and unidirectional devices. \textit{Opt. Express} \textbf{17}, 5265-5272 (2009).
\bibitem{012}Lepri, S. and Casati, G. Asymmetric Wave Propagation in Nonlinear Systems. \textit{Phys. Rev Lett.} \textbf{106}, 164101(2011).
 \bibitem{013} Lepri, S. and Casati, G. Nonreciprocal wave propagation through open, discrete nonlinear Schr\"{o}dinger dimers. In \textit{Localized Excitations in Nonlinear Complex Systems: Current State of the Art and Future Perspectives}. Nonlinear Systems and Complexity, Vol. \textbf{7} (Springer, Cham,
Switzerland, 2014). arXiv:1211.4996.
 \bibitem{014} Kosevich, Y. A. Fluctuation subharmonic and multiharmonic phonon transmission and Kapitza conductance between crystals with very different vibrational spectra. \textit{Phys. Rev. B} \textbf{52}, 1017 (1995).
\bibitem{015} Scalora, M., Dowling, J. P., Bowden, C. M., and Bloemer, M. J. The photonic band edge optical diode. \textit{J. Appl. Phys.} \textbf{76}, (1994).
 \bibitem{016} Tocci, M. D., Bloemer, M. J., Scalora, M., Dowling, J. P., and Bowden, C. M. Thin-film nonlinear optical diode. \textit{Appl. Phys. Lett.} \textbf{66}, (1995).
 \bibitem{017} Terraneo, M., Peyrard, M., and Casati, G. Controlling the Energy Flow in Nonlinear Lattices: A Model for a Thermal Rectifier. \textit{Phys. Rev. Lett.} \textbf{88}, 094302 (2002).
 \bibitem{018} Segal, D. and Nitzan, A. Spin-Boson Thermal Rectifier. \textit{Phys. Rev. Lett.} \textbf{94}, 034301 (2005).
\bibitem{019} Liang, B., Yuan, B., and Cheng, J. C. Acoustic Diode: Rectification of Acoustic Energy Flux in One-Dimensional Systems. \textit{Phys. Rev. Lett.} \textbf{103}, 104301 (2009).
\bibitem{020} Campbell, D. K., Flach, S., and Kivshar, Y. S. Localizing Energy Through Nonlinearity and Discreteness. \textit{Physics Today} \textbf{57}, 43 (2004).
\bibitem{021} Lederer, F., Stegeman, G. I., Christodoulides, D. N., Assanto, G., Segev, M., and Silberberg, Y. Discrete solitons in
optics. \textit{Phys. Rep.} \textbf{463}, 1-126 (2008).
\bibitem{022} Flach, S. and Gorbach, A. V. Discrete Breathers. \textit{Phys. Rep.} \textbf{467}, 1-116 (2008).
\bibitem{023} Wasay, M. A. Nonreciprocal wave transmission through an extended discrete nonlinear Schr\"{o}dinger dimer. \textit{Phys. Rev. E} \textbf{96}, 052218 (2017).
\bibitem{024} Mej\'{i}a-Cort\'{e}s, C., Vicencio, R. A. and Malomed, B. A. Mobility of solitons in one-dimensional lattices with the cubic-quintic nonlinearity. \textit{Phys. Rev. E} \textbf{88}, 052901 (2013).
\bibitem{025} Carretero-Gonz\'{a}lez, R., Talley, J. D., Chong, C. and Malomed, B. A. Multistable Solitons in the Cubic-Quintic Discrete Nonlinear Schr\"{o}dinger Equation. \textit{Physica D} \textbf{216}, 77-89 (2006).
\bibitem{26}  Bai, X. D., Malomed, B. A. and Deng, F. G. Unidirectional transport of wave packets through tilted discrete breathers in nonlinear lattices with asymmetric defects. \textit{Phys. Rev. E} \textbf{94}, 032216 (2016).
\bibitem{27}  Bai, X. D. and Xue, J. K. Discrete breather and its stability in a general discrete nonlinear Schrödinger equation with disorder. \textit{Phys. Rev. E} \textbf{86}, 066605 (2012).
\bibitem{28}  Bai, X. D., Ai, Q., Zhang, M., Xiong, J., Yang, G. J. and Deng, F. G.  Stability and phase transition of localized modes in Bose-Einstein condensates with both two and three-body interactions. Ann. Phys. \textbf{360}, 679-693, (2015).
\bibitem{026} Tsironis, G. and Hennig, D. Wave transmission in nonlinear lattices. \textit{Phys. Rep.} \textbf{307}, 333-342 (1999).
\bibitem{027} Delyon, F., L\'{e}vy, Y. and Souillard, B. Nonperturbative Bistability in Periodic Nonlinear Media. \textit{Phys. Rev. Lett.} \textbf{57}, 2010 (1986).
\bibitem{028} Li, Q., Chan, C. T., Ho, K. M. and Soukoulis, C. M. Wave propagation in nonlinear photonic band-gap materials. \textit{Phys. Rev. B.} \textbf{53}, 15577 (1996).
\bibitem{029} \'{D}Ambroise, J., Kevrekidis, P. G. and Lepri, S. Asymmetric wave propagation through nonlinear $\mathcal{PT}$-symmetric oligomers. \textit{J. Phys. A: Math. Theor.} \textbf{45}, 444012 (2012).
\bibitem{030} Johansson, E. Model of a Wave Diode in a Nonlinear System (Dissertation) (2014). Retrieved from http://urn.kb.se/resolve?urn=urn:nbn:se:liu:diva-111236
\bibitem{031} Law, D., \'{D}Ambroise, J., Kevrekidis, P. G. and Kip, D. Asymmetric Wave Propagation Through Saturable Nonlinear Oligomers. \textit{Photonics} \textbf{1}, 390-403 (2014).

\end{thebibliography}
\end{document}